\documentstyle[psfig]{mn}

\def\etal{{\it et al.\ }}
\def\eg{{\it e.g.\ }}

\def\ie{{\it i.e.\ }}

\def\spose#1{\hbox to 0pt{#1\hss}}
\def\approxlt{\mathrel{\spose{\lower 3pt\hbox{$\sim$}}
	\raise 2.0pt\hbox{$<$}}}
\def\approxgt{\mathrel{\spose{\lower 3pt\hbox{$\sim$}}
	\raise 2.0pt\hbox{$>$}}}
\def\approxpropto{\mathrel{\spose{\lower 3pt\hbox{$\sim$}}
	\raise 2.0pt\hbox{$\propto$}}}
\mathchardef\twiddle="2218

\def\multleft#1{\hbox to size{\vbox {\halign {\lft{##}\cr #1}}\hfill}\par}
\def\multright#1{\hbox to size{\vbox {\halign {\rt{##}\cr #1}}\hfill}\par}

\def\today{\ifcase\month\or January\or February\or March\or April\or May\or
      June\or July\or August\or September\or October\or November\or December\fi
      \space\number\day, \number\year}
\def\<{\thinspace}

\def\apc{\rm atom cm$^{-2}$}

\def\cm{{\rm\thinspace cm}}
\def\erg{{\rm\thinspace erg}}

\def\km{{\rm\thinspace km}}

\def\Mpc{{\rm\thinspace Mpc}}
\def\Msun{\hbox{$\rm\thinspace M_{\odot}$}}

\def\s{{\rm\thinspace s}}
\def\yr{{\rm\thinspace yr}}

\def\ergpcmsqps{\hbox{$\erg\cm^{-2}\s^{-1}\,$}}

\def\ergps{\hbox{$\erg\s^{-1}\,$}}

\def\kmps{\hbox{$\km\s^{-1}\,$}}

\def\Msunpyr{\hbox{$\Msun\yr^{-1}\,$}}

\def\kmpspMpc{\hbox{$\kmps\Mpc^{-1}$}}

\def\apc{\rm atom cm$^{-2}$}

\title[Chandra observations of RXJ1347.5-1145]
{Chandra observations of RXJ1347.5-1145: the distribution of mass in 
the most X-ray luminous galaxy cluster known}
\author[S.W. Allen, R.W. Schmidt and A.C. Fabian]
{\parbox[]{6.in} {S.W. Allen, R.W. Schmidt and A.C. Fabian \\
\footnotesize
Institute of Astronomy, Madingley Road, Cambridge CB3 0HA
}}
\begin{document}
\maketitle
\begin{abstract}
We present Chandra observations of RXJ1347.5-1145, the most X-ray
luminous cluster of galaxies known. We report the discovery of a
region of relatively hot, bright X-ray emission, located approximately
20 arcsec to the southeast of the main X-ray peak, at a position
consistent with the region of enhanced Sunyaev-Zeldovich effect
reported by Komatsu \etal (2001). We suggest that this region contains
shocked gas resulting from a recent subcluster merger event.
Excluding the data for the southeast quadrant, the cluster appears
relatively relaxed. The X-ray gas temperature rises from  $kT \sim 6$
keV within the central  $25\,h_{50}^{-1}$ kpc radius to a mean value
of $\sim 16$ keV between $0.1-0.5\,h_{50}^{-1}$ Mpc. The mass profile
for the relaxed regions of the cluster, determined under the
assumption of  hydrostatic equilibrium, can be parameterized by a
Navarro,  Frenk \& White (1997) model with a scale radius  $r_{\rm s}
\sim 0.4\,h_{50}^{-1}$ Mpc and a concentration  parameter $c \sim
6$. The best-fit Chandra mass model is in good  agreement with
independent measurements from weak  gravitational lensing studies.
Strong lensing data for the central regions of the cluster can be also
explained by the introduction of an additional mass clump  centred on
the second brightest galaxy. We argue that this galaxy is likely to
have been the dominant galaxy of the recently merged subcluster.
\end{abstract}

\begin{keywords}
galaxies: clusters: individual: RXJ1347.5-1145 -- gravitational lensing --
X-rays: galaxies -- cooling flows  -- intergalactic medium
\end{keywords}

\section{Introduction}

The launch of the Chandra Observatory (Weisskopf \etal 2000)  in 1999
July has provided the first opportunity for detailed,
spatially-resolved X-ray spectroscopy of clusters of galaxies  at
moderate to high redshifts. The Advanced CCD Imaging Spectrometer
(ACIS) on Chandra permits direct, simultaneous measurements of the X-ray gas
temperature and density profiles in clusters and, via the hydrostatic
assumption, the mass distributions, spanning scales from $r \sim 10$
kpc  in cluster cores out to $r \sim 1$ Mpc at the detection limits.

In this paper, we present the first results from Chandra observations
of  RXJ1347.5-1145, the most X-ray luminous galaxy cluster known
(Schindler \etal 1995). This cluster  has been the subject of several
previous X-ray, optical and Sunyaev-Zeldovich (SZ) effect studies (\eg
Schindler \etal 1995, 1997; Fischer \& Tyson 1997, Allen 1998;
Komatsu \etal 1999, 2001; Pointecouteau \etal 1999, 2001). Here, we
present the first measurements of the X-ray temperature structure
within the cluster and place tight  constraints on the total  mass
distribution. We compare our results with those from 
strong and weak lensing analyses.  A comparison of the Chandra and SZ
results, and a determination of the  Hubble Constant from the combined
data sets, is presented elsewhere (Schmidt \& Allen 2002, in
preparation).

Except where stated otherwise, the cosmological parameters  $H_0$=50
\kmpspMpc, $\Omega = 1$ and  $\Lambda = 0$ are assumed. At
the redshift of RXJ1347.5-1145  ($z=0.451$), an angular scale of 1
arcsec corresponds to physical size  of  6.81 kpc in this cosmology.

\begin{figure*}
\vspace{0.5cm}
\hbox{
\hspace{0.2cm}\psfig{figure=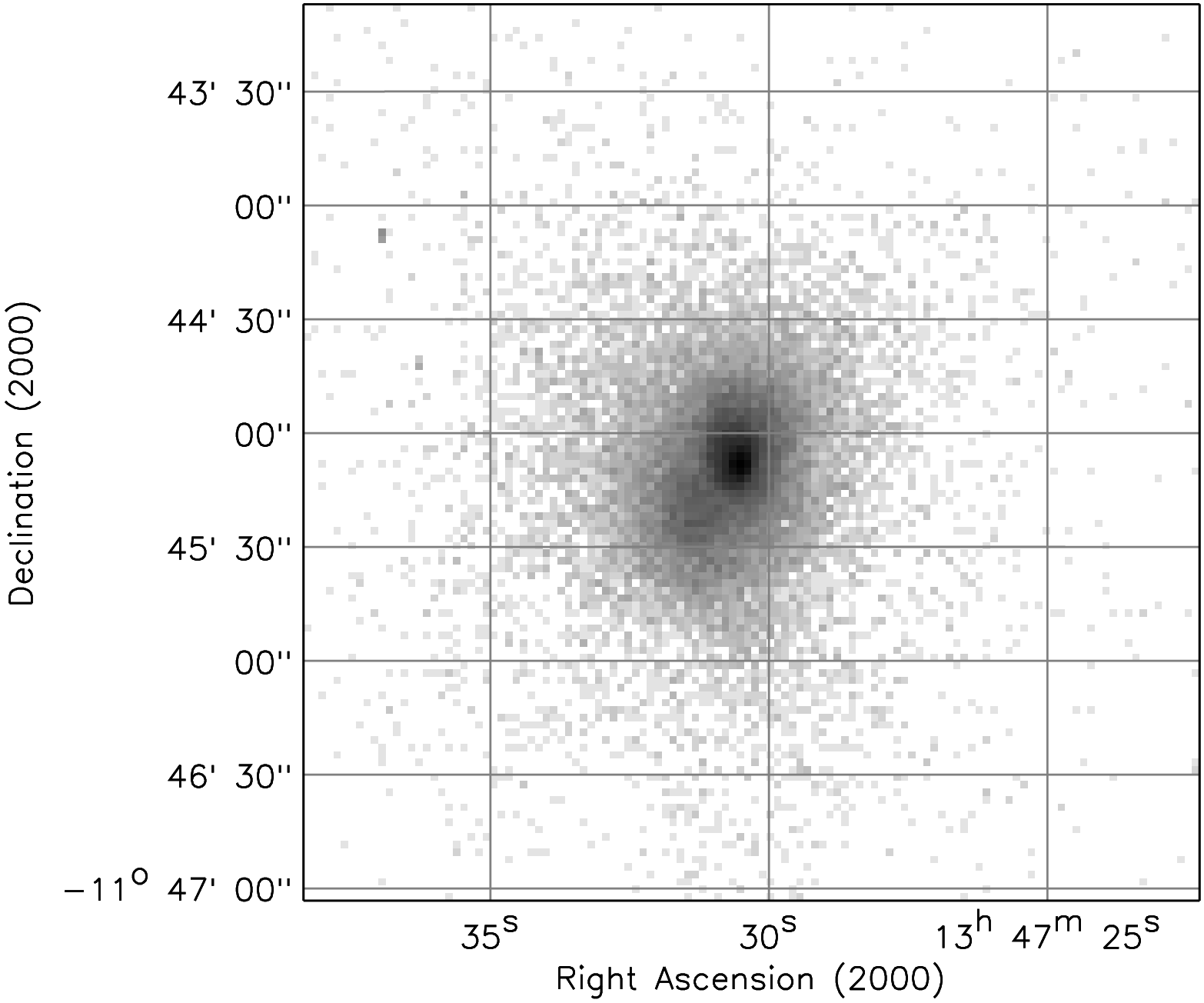,width=0.45 \textwidth,angle=0}
\hspace{0.8cm}\psfig{figure=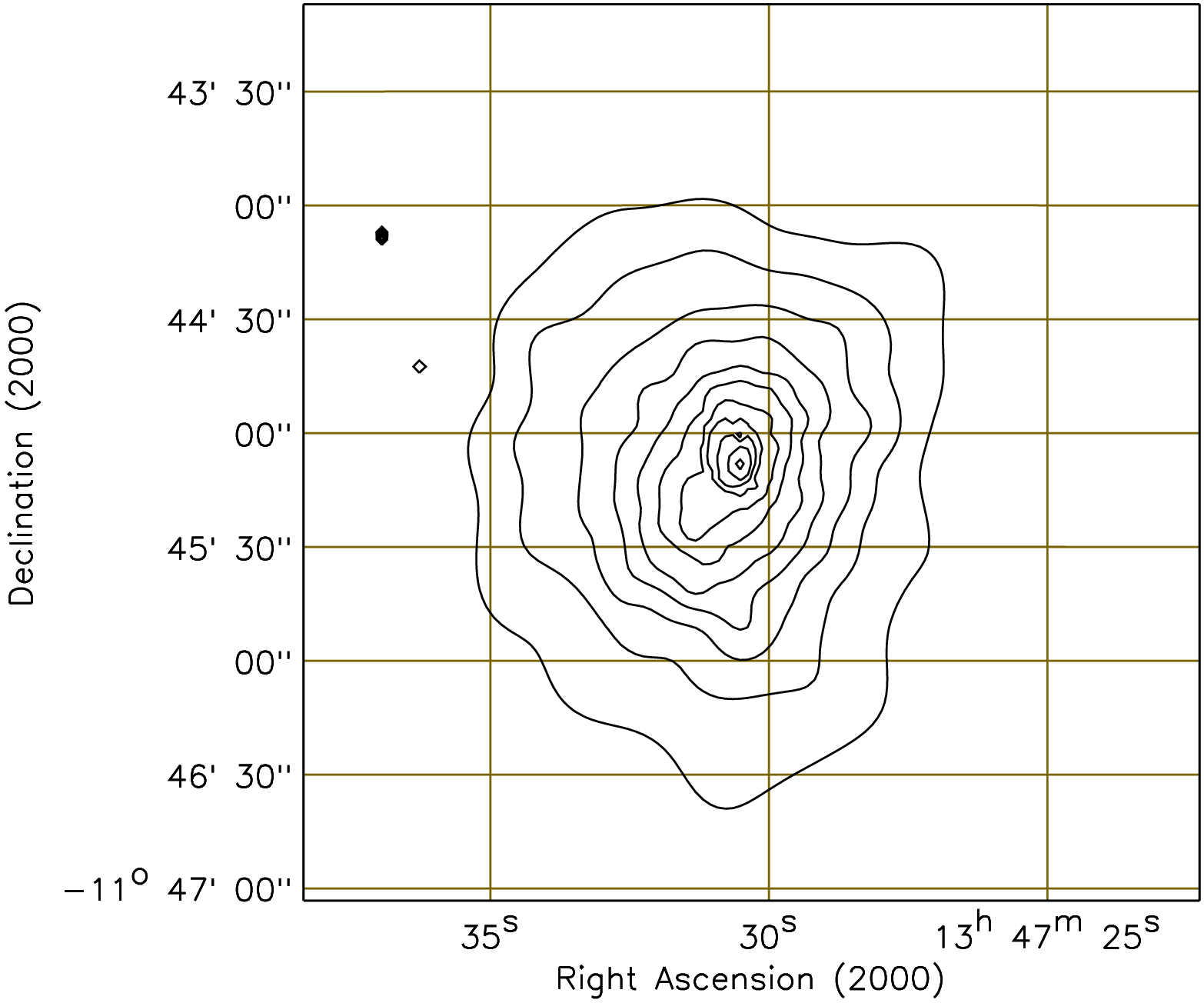,width=0.45 \textwidth,angle=0}
}
\caption{(a) The raw $0.3-7.0$ keV Chandra image of RXJ1347.5-1145. 
The pixel size is 4 detector pixels (1.97 arcsec). (b) Contour 
plot of the same region, adaptively smoothed using the code of Ebeling \etal 
(2002) with a threshold value of $3\sigma$. The 
contours have equal logarithmic spacing. }\label{fig:im1} 
\end{figure*}

\section{Observations}

The  Chandra observations of  RXJ1347.5-1145 were carried out using
the  ACIS on 2000 March 05 and 2000 April 29. The target was observed
in the  back-illuminated S3 detector and positioned near the centre of
node-1  on CCD 7. The light curves for both observations were of high
quality with  no strong background flares. The net good exposure times
were 8.9 and  10.0 ks, respectively, giving a total good exposure time
of 18.9 ks. The  focal plane temperature at the time of both
observations was -120C.

We have used the  CIAO software and the level-2 events files  provided
by the standard Chandra pipeline processing for our analysis.  Only
those X-ray events with  grade classifications of 0,2,3,4 and 6 were
included in our final cleaned data sets.

\section{X-ray imaging}

\subsection{X-ray morphology}

The raw $0.3-7.0$ keV image of the central $6 \times 6$ arcmin$^2$
region  of the cluster, from the combined 18.9 ks data set, is shown
in Fig.  \ref{fig:im1}(a). The pixel size is $1.97  \times 1.97$
arcsec$^2$,  corresponding to $4 \times 4$ raw detector
pixels. Fig. \ref{fig:im1}(b)  shows an adaptively smoothed contour
plot of the same data, using the  smoothing algorithm of Ebeling,
White \& Rangarajan (2002). The images  reveal a number of notable
features. Firstly, we see a  sharp central surface brightness peak at
a position 13h47m30.63 -11d45m09.3s (J2000.), in good agreement with the
optical centroid for the dominant cluster galaxy,  13h47m30.5
-11d45m09s (J2000.; Schindler \etal 1995). Secondly, we   identify a
region of enhanced emission $\sim 20$ arcsec to the  southeast of the
X-ray peak (see below). Thirdly, on large scales  ($r \sim 80$ arcsec)
we detect an extension of the X-ray isophotes to the  south.
 
\begin{figure}
\vspace{0.5cm}
\hbox{
\hspace{0.6cm}\psfig{figure=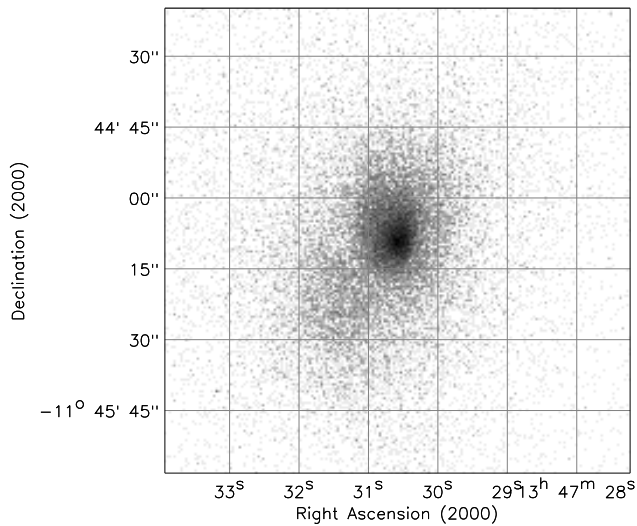,width=0.45 \textwidth,angle=0}
}
\caption{A raw $0.3-7.0$ keV image of the central regions of 
RXJ1347.5-1145 on a finer spatial scale (pixel size $0.492 \times 0.492$ 
arcsec$^2$, equivalent to $1\times1$ raw detector pixels).}\label{fig:im2} 
\end{figure}

Fig. \ref{fig:im2} shows the data for the central regions of the
cluster on  a finer $0.492  \times 0.492$ arcsec$^2$ scale
(corresponding to $1  \times 1$ raw  detector pixels). From this
figure it appears that the region of  enhanced emission to the
southeast of the X-ray peak  has a roughly circular morphology with a
flattened northwest edge.  Beyond this edge, at a radius $r \sim 10$
arcsec, the X-ray subclump appears  to be separated from the main cluster
core by a  valley of reduced  emission. The spectral analysis in
Section 4.2 shows that the X-ray  subclump has a hotter temperature
than the gas at the  same radius in other directions from the X-ray
peak  and probably contains shocked gas resulting from a recent 
subcluster  merger event.

Komatsu \etal (2001) report the detection of a region of enhanced SZ
effect at a position coincident with the X-ray subclump. The enhanced
SZ effect is consistent with the higher temperature and density for 
this region measured from the Chandra data. 

No strong X-ray point sources are detected in the Chandra S3 field. (The
brightest point source has a flux of $3.1\times10^{-14}$ \ergpcmsqps 
in the $0.5-7.0$ keV band; Gandhi \etal, in preparation.)

\subsection{Surface brightness profiles}

\begin{figure*}
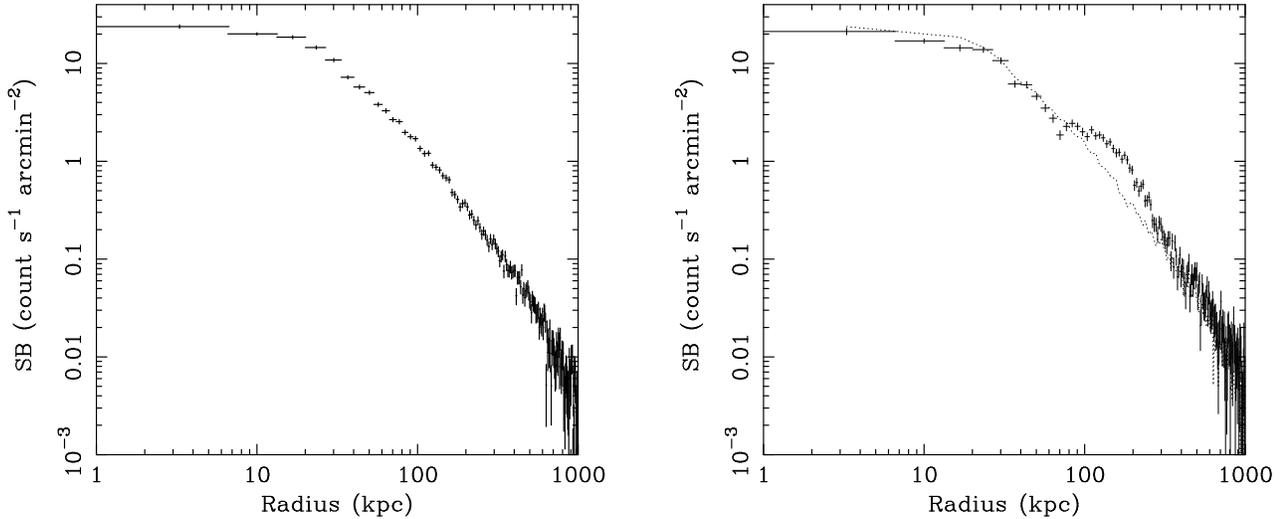

\vspace{0.5cm}
\hbox{
\hspace{0.2cm}\psfig{figure=fig3a.ps,width=0.45 \textwidth,angle=270}
\hspace{0.8cm}\psfig{figure=fig3b.ps,width=0.45 \textwidth,angle=270}}
\caption{Background-subtracted, flat-fielded, azimuthally-averaged 
radial surface brightness profiles for RXJ1347.5-1145 in the $0.3-7.0$ keV 
band. The bin-size is 0.984 arcsec (6.70 kpc). (a) The results for position 
angles of $180-90$ degrees. The profile appears regular and relaxed (b) 
The results for position angles of $90-180$ degrees (the southeast 
quadrant). The dotted curve shows the results for the $180-90$ degree region. 
Note the clear excess of emission in the southeast quadrant between radii 
of $10-45$ arcsec ($70-300$ kpc).}\label{fig:surbri}
\end{figure*}

The azimuthally-averaged, $0.3-7.0$ keV X-ray surface brightness
profile  for RXJ1347.5-1145 for position angles of $180-90$ degrees
(i.e. excluding  the southeast quadrant) and for the disturbed,
southeast quadrant are  shown in Figs. \ref{fig:surbri}(a,b),
respectively. The profiles have been  flat-fielded and  background
subtracted using an on-chip region free from cluster emission.  The
bin-size is 2 detector pixels (0.984 arcsec).

We see that the data for position angles of $180-90$ degrees appear
regular  and, for $r < 500$ kpc, can be approximated ($\chi^2 = 102$
for 72 degrees  of freedom) by a $\beta$-model (\eg Jones \& Forman
1984) of the form  $S_{\rm X}(r) = S(0)\left [ {1+{(r/r_{\rm c}})^2}
\right ]^{1/2-3\beta}$  with a core radius $r_{\rm c} = 29.2\pm0.7$
kpc and a slope parameter  $\beta =0.535\pm0.003$ ($1 \sigma$
errors). Between  radii of $0.1-1.0$ Mpc we find evidence for a
steepening of the surface  brightness profile with increasing radius:
for $0.1 < r < 1.0$ Mpc the  data can be described ($\chi^2 = 135$ for
131 degrees of freedom) by a  simple broken power-law model, with a
break at a radius of $487^{+23}_{-16}$  kpc, and slopes in the regions
internal and external to the break radius of  $-2.25\pm0.03$ and
$-3.66 \pm 0.18$, respectively. A similar result was obtained for
Abell 2390 by Allen, Ettori \& Fabian (2001b), although in that case 
the slope interior to the break radius was flatter.

The data for the southeast quadrant containing the X-ray subclump
reveal a clear excess of emission relative to other
directions. Enhanced  emission is detected between radii of $\sim
70-300$ kpc, with a maximum  enhancement of a factor $\sim 3$ at a
radius $r \sim 180$ kpc.

\section{Spatially-resolved spectroscopy}

\subsection{Method}

For our spectral analysis, we have separated the southeast quadrant
containing the X-ray subclump from the rest of the cluster.  The data
for position angles of $180-90$ (the relaxed part of the cluster) were
divided into the annular regions detailed in Table 1.\footnote{For the
southeast quadrant, we have extracted a single spectrum covering radii
$60-195$ kpc (18-58 pixels), where the emission enhancement in Fig.~
\ref{fig:im2}  is most obvious. } A single spectrum was  extracted for
each region in 1024 Pulse Invariant (PI) channels. The  spectra were
re-grouped to contain a minimum of 20 counts per PI channel thereby
allowing $\chi^2$ statistics to be used.

Background spectra, appropriate  for the regions studied, were
extracted from the  ACIS-S3 blank-field data sets provided by Maxim
Markevitch  (which are available from the Chandra X-ray Center).
Separate photon-weighted response matrices and  effective area files
were constructed for each region  using the available calibration and
response files appropriate for the -120C focal plane
temperature.\footnote{For each $32 \times 32$ pixel$^2$ sub-region of
the  S3 chip, a spectral response (.rmf) and an auxiliary response
(.arf) matrix  were created using the CIAO tools $mkrmf$ and $mkarf$,
respectively. For each  of the annular regions studied, the number of
source counts in  each $32 \times 32$ pixel$^2$ sub-region was
determined.  The individual .rmf and .arf files were then combined
using the FTOOLS  programs $addrmf$ and $addarf$ (using a script
provided by Roderick  Johnstone) to form a counts-weighted spectral
response and auxiliary  response matrix appropriate for each annulus.}

The data from both Chandra observations were modelled simultaneously
using the XSPEC code (version 11.01; Arnaud 1996). We have limited our
analysis to the $0.5-7.0$ keV energy band, over which the calibration
of the back-illuminated CCD detectors is currently best understood.

The spectra have been modelled using the MEKAL  plasma emission code of
Kaastra \& Mewe (1993; incorporating the Fe L calculations of Liedhal,
Osterheld \& Goldstein 1995) and the photoelectric absorption models
of  Balucinska-Church \& McCammon (1992). We have fitted each annular
spectrum using a simple, single-temperature model with the absorbing
column density fixed at the nominal Galactic value ($N_{\rm H} = 4.85
\times 10^{20}$\apc; Dickey \& Lockman 1990). The free parameters in
this  model were the temperature ($kT$), metallicity ($Z$; measured
relative to  the solar photospheric values of Anders  \& Grevesse
1989, with the various elements assumed to be present  in their solar
ratios) and emission measure. We note that including the absorbing
column density as an additional free parameter did not result in 
significant improvements in the goodness of fit.

\begin{table}
\begin{center}
\caption{The results from the analysis of the annular spectra
(covering position angles $180-90$ degrees).  Temperatures ($kT$) are
in keV and metallicities ($Z$) in solar units.  The absorbing column
density has been  fixed at the nominal Galactic value of $4.85 \times
10^{20}$\apc. The total $\chi^2$ values  and number of degrees of
freedom (DOF) in the fits are listed in column 4.  Error bars are the
$1\sigma$ ($\Delta \chi^2=1.0$)  confidence limits on a single
interesting parameter.  }
\begin{tabular}{ c c c c c }
&&&&  \\                                                                                                                   \hline
Radius (kpc)     & ~ &  $kT$                    &    $Z$                 &   $\chi^2$/DOF \\

\hline
$0-23.5$     & ~ &  $7.04^{+0.79}_{-0.60}$  & $0.48^{+0.17}_{-0.17}$  &127.6/101   \\
$23.5-50.3$    & ~ &  $9.03^{+0.99}_{-0.74}$  & $0.66^{+0.16}_{-0.15}$  &126.4/130   \\
$50.3-97.2$   & ~ & $11.23^{+1.22}_{-1.02}$  & $0.60^{+0.16}_{-0.15}$  &190.6/164   \\
$97.2-147$  & ~ & $13.82^{+2.09}_{-1.69}$  & $0.07^{+0.17}_{-0.07}$  &106.3/119   \\
$147-245$  & ~ & $19.37^{+4.07}_{-2.83}$  & $0.28^{+0.26}_{-0.26}$  &134.7/134   \\
$245-345$  & ~ & $13.51^{+2.73}_{-2.11}$  & $0.81^{+0.34}_{-0.30}$  & 78.2/89   \\
$345-492$  & ~ & $18.08^{+5.84}_{-4.23}$  & $0.79^{+0.47}_{-0.42}$  & 92.6/85   \\
$492-737$  & ~ & $12.40^{+4.52}_{-2.61}$  & $0.44^{+0.33}_{-0.35}$  & 80.2/81   \\
&&&&  \\
$0-97.2$    & ~ & $9.32^{+0.57}_{-0.52}$   & $0.55^{+0.09}_{-0.09}$  & 457.9/399   \\
$97.2-737$  & ~ & $15.48^{+1.53}_{-1.16}$  & $0.34^{+0.12}_{-0.12}$  & 502.7/516   \\
$0-737$    & ~ & $12.00^{+0.62}_{-0.59}$  & $0.41^{+0.07}_{-0.07}$  & 987.8/917   \\
&&&& \\
\hline
\end{tabular}
\end{center}
\end{table}

\subsection{Results}

\begin{figure}
\vspace{0.5cm}
\hbox{
\hspace{0.0cm}\psfig{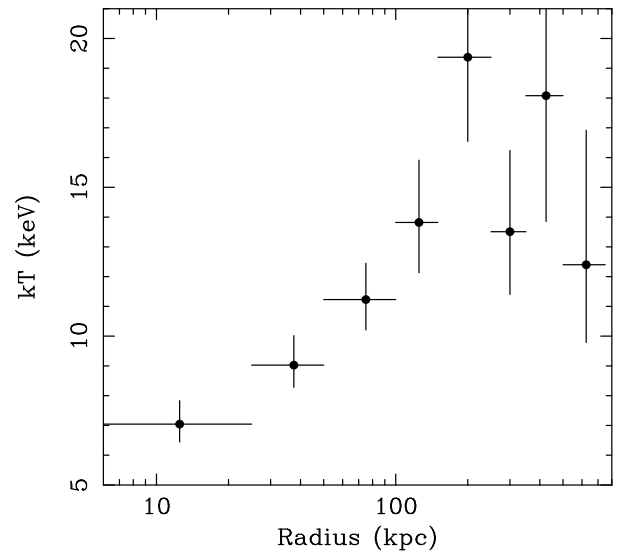}
}
\caption{The projected X-ray gas temperature profile 
(and $1\sigma$ errors) measured from the Chandra data in the 
$0.5-7.0$ keV energy band. }\label{fig:kt}
\end{figure}

The best-fit parameter values and $1\sigma$ ($\Delta \chi^2 = 1.0$)
confidence limits determined from the fits to the annular spectra
(excluding the southeast quadrant) in the $0.5-7.0$ keV band  are
summarized in Table 1. The projected temperature  profile determined
with this model is shown in Fig. \ref{fig:kt}.  The temperature rises
from a mean value of $7.0^{+0.8}_{-0.6}$ keV  within $r=24$ kpc to
$kT=15.8^{+1.6}_{-1.2}$ keV over the  $0.1-0.5$ Mpc region.

A fit with the same model to the data for the southeast quadrant
between  radii of $60-195$ kpc (where the X-ray enhancement is most obvious 
in the image) gives a best-fit temperature  $kT =
18.0^{+2.7}_{-2.3}$ keV. (A similar result, $kT =  18.2^{+3.8}_{-2.9}$
keV, is obtained using a circular region of radius 10 arcsec centred on
the enhancement.)  In other directions, the mean temperature  between
radii of  $60-195$ kpc is $kT = 12.7\pm1$ keV. The higher  temperature
and enhanced X-ray surface  brightness in the southeast quadrant are
consistent with the X-ray  gas in that region having undergone shock
compression (Section 8).

We find marginal evidence for a metallicity gradient in the cluster,
with a mean metallicity for the central 100 kpc radius of
$Z=0.55\pm0.09$ solar,  as opposed to $Z=0.34\pm0.12$ solar for the
$100-740$ kpc region.

\subsection{Spectral deprojection analysis}

The results discussed above are based on the analysis of projected
spectra. We have also  carried out a simple deprojection analysis of
the Chandra  spectral data using the method described by Allen \etal
(2001b).

For this analysis we have used the same annular regions (covering
position angles of $180-90$ degrees) and have assumed that the
emission  from each spherical shell (the shells are defined by the
same inner and  outer radii as the annular regions) is isothermal and
absorbed by the  Galactic column density.  The fit to the outermost
annulus is used to determine the temperature and emission measure in
the  outermost spherical shell. The contribution from that shell to
each inner  annulus is then determined by purely geometric factors
(\eg Kriss, Cioffi \&  Canizares 1983). The fit to the second annulus
inward is used  to determine the parameters for the second spherical
shell, and so forth,  working inwards.

\begin{figure}
\vspace{0.5cm}
\hbox{
\hspace{0.0cm}\psfig{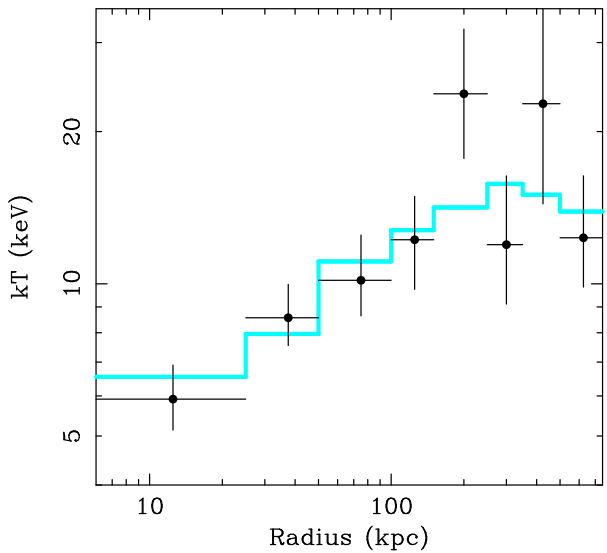}
}
\caption{The predicted deprojected temperature profile (grey curve)
determined  from 100 Monte-Carlo simulations using the best-fitting
NFW mass model  (with $r_{\rm s} = 0.40$ Mpc, $c=5.87$ and $\sigma =
1450$\kmps; Section 5).  The predicted profile has been binned to the
same spatial  resolution as the spectral deprojection results (solid
points; Section 4.3),  which have been overlaid. The agreement between
the deprojected spectral  results and the best-fit NFW mass model
predictions (reduced $\chi^2_{\nu} = 0.87$  for $\nu=6$ degrees of
freedom) indicates that the NFW mass model provides a  good
description of the spatially-resolved Chandra
spectra.}\label{fig:kt_xspec3d}
\end{figure}

The data for all eight annular spectra were fitted simultaneously in
order to determine the parameter values and confidence
limits.  The metallicity was linked to take the same value at all
radii. The temperature profile determined with the spectral
deprojection  method is shown in Fig. \ref{fig:kt_xspec3d}.

\subsection{Comparison with previous work} 

The mean emission-weighted temperature and metallicity for
RXJ1347.5-1145 of $kT = 12.2\pm0.6$ keV and $Z=0.42\pm0.07$ solar,
respectively, measured in the $0.0-0.74$ Mpc range over the full  360
degrees, are in good agreement with the values reported by  Allen \&
Fabian (1998; $kT = 12.5^{+0.9}_{-0.8}$ keV,  $Z=0.38^{+0.11}_{-0.10}$
solar)  using ASCA data and a similar spectral model. Excluding the
data for  the southeast quadrant, we obtain only small changes in
these results:  $kT = 12.0\pm0.6$ keV and $Z=0.41\pm0.07$ solar.
The Chandra results are also  in good agreement with those
reported by Schindler \etal (1997) using the same ASCA data and a 3
arcmin radius  aperture ($kT = 11.8^{+1.6}_{-1.0}$ keV,  $Z=
0.39^{+0.12}_{-0.13}$ solar). The emission-weighted temperature
measured with Chandra is slightly  cooler than the value of
$14.5^{+1.7}_{-1.5}$ keV measured with BeppoSAX  by Ettori, Allen \&
Fabian (2001), as can be expected given the observed temperature profile
(Figs. \ref{fig:kt},\ref{fig:kt_xspec3d}) and the different response
characteristics of the instruments. (The data for the
southeast quadrant are not excluded from the BeppoSAX or ASCA
studies.)

The `ambient' emission-weighted temperature measured with Chandra in
the  $0.1-0.5$ Mpc range ({\it i.e.} excluding the data for the 
southeast quadrant, the central, cool region, and the regions at large 
radii where the temperature may  start
to decline again) of $kT=15.8^{+1.6}_{-1.2}$ keV is in good agreement
with the value of $15.9^{+6.5}_{-2.6}$ keV estimated  from BeppoSAX
data by Ettori  \etal (2001) using a spectral model  which included a
constant-pressure cooling flow component. The Chandra result  is also
consistent with the ambient temperature of $kT = 26^{+8}_{-12}$ keV
estimated from ASCA data using a similar constant-pressure cooling
flow model (Allen \& Fabian 1998), and $kT = 18.6^{+4.1}_{-3.0}$ keV
from a re-analysis of  the ASCA data by the same authors using the more 
appropriate `isothermal' cooling flow  models of Nulsen (1998).

The emission weighted ambient temperature determined from the
Chandra data in the $0.1-0.5$ Mpc region is in good agreement with
the mean gas mass-weighted temperature of $16.1^{+5.3}_{-2.7}$
keV  measured within $r_{2500}$ ($r=0.72$ Mpc) by Allen, Schmidt \&
Fabian (2001c).  We note that this temperature is also consistent with
the predicted value of $kT = 16.8$  keV, using the cooling-flow
corrected $kT_{\rm X}/L_{\rm Bol}$ relation of  Allen \& Fabian (1998;
we assume a bolometric luminosity of $\sim 2.2  \times 10^{46}$
\ergps~as measured by ASCA).

\section{Measurement of the cluster mass profile}

\begin{figure}
\vspace{0.5cm}
\hbox{
\hspace{0.0cm}\psfig{figure=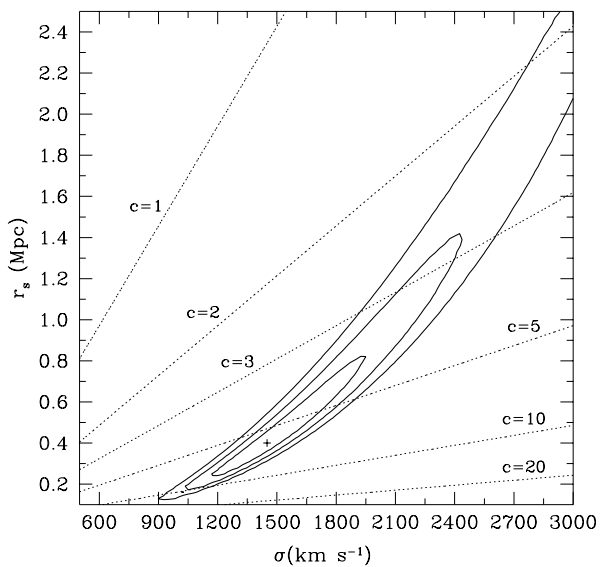,width=0.45 \textwidth,angle=0}}
\caption{Contour plot showing the 68.3\% ($1\sigma$), 95.4\%
($2\sigma$) and 99.73\% ($3\sigma$) confidence contours on the scale
radius, $r_{\rm s}$, and the effective velocity dispersion, $\sigma$,
for  the NFW mass models.   Contours of constant concentration
parameter, $c$, are marked with dashed  lines. The best-fit model is
marked by a plus sign.}\label{fig:contour}
\end{figure}

\subsection{Method}

The observed X-ray surface brightness profile (Fig. \ref{fig:surbri}a)
and deprojected, spectrally-determined temperature profile
(Fig. \ref{fig:kt_xspec3d}) may together be used to determine the
X-ray gas  mass and total mass profiles in the cluster. For this
analysis, we have used an enhanced version of the image deprojection
code  described by White, Jones \& Forman (1997), and have followed
the methods outlined by Allen \etal (2001b) and Schmidt \etal
(2001).

A variety of simple parameterizations for the cluster mass
distribution were examined to establish which could provide an
adequate description of the Chandra data (see below). The  best-fit
parameter values and confidence limits were determined by examining
parameter grids and evaluating $\chi^2$ for  each set of
parameters.\footnote{The observed surface brightness  profile
(Fig.~\ref{fig:surbri}a) and a particular parameterized mass  model
are together used to predict the temperature profile of the X-ray
gas. (We use the median temperature  profile determined from 100
Monte-Carlo simulations in which the surface brightness profile is
perturbed according to the  statistical uncertainties at each
radius. The outermost pressure is fixed using an iterative technique
which ensures a smooth pressure gradient in these regions.)  The
predicted temperature profile is rebinned to the same binning as the
spectral results and the $\chi^2$  difference between the observed and
predicted, deprojected temperature  profiles is calculated. The
parameters for the mass model are  stepped through a regular grid of
values in the $r_{\rm s}$-$\sigma$ plane (see text) to determine the
best-fit values and 68 per cent confidence limits. The gas mass
profile is determined to high precision at each grid point using the
observed surface brightness profile and model temperature profile.}
Spherical symmetry and hydrostatic equilibrium are assumed throughout.

\subsection{NFW mass models}
\label{nfwmodels}

We find that a good fit ($\chi^2_{\rm min} = 5.2$ for 6 degrees of freedom, 
hereafter DOF) to the Chandra data for RXJ1347.5-1145 can be obtained using a 
Navarro, Frenk \& White (1997, hereafter NFW) model:

\begin{equation}
\rho(r) = {{\rho_{\rm crit}(z) \delta_{\rm c}} \over {  ({r/r_{\rm s}}) 
\left(1+{r/r_{\rm s}} \right)^2}},
\end{equation}

\noindent where $\rho(r)$ is the mass density, $\rho_{\rm crit}(z) =
3H(z)^2/ 8 \pi G$ is  the critical density for closure and

\begin{equation}
\delta_{\rm c} = {200 \over 3} { c^3 \over \left[ {{\rm ln}(1+c)-{c/(1+c)}}\right]}.
\end{equation}

\noindent The best-fit scale radius, $r_{\rm s} =
0.40^{+0.24}_{-0.12}$ Mpc and  the concentration parameter,
$c=5.87^{+1.35}_{-1.44}$ (68 per cent confidence  limits). The
normalization of the mass profile may also be expressed in terms  of
an effective velocity dispersion, $\sigma = \sqrt{50} H(z) r_{\rm s}
c = 1450^{+300}_{-200}$\kmps (with $r_{\rm s}$ in units of Mpc).  The
deprojected X-ray gas temperature profile predicted by the
best-fitting  NFW mass model (given the observed surface brightness
profile) is shown  overlaid on the observed spectral results in
Fig. \ref{fig:kt_xspec3d}.

\begin{figure}
\vspace{0.5cm}
\hbox{
\hspace{0.0cm}\psfig{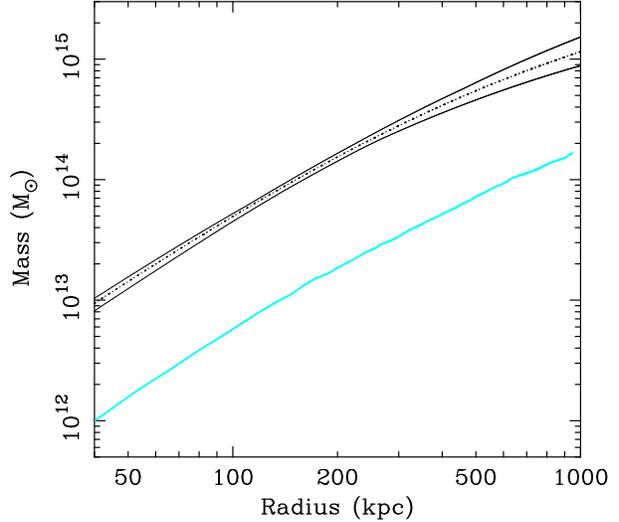}}
\caption{The integrated (3-dimensional) mass profile and $1\sigma$
errors for RXJ1347.5-1145 determined from the Chandra data using the
NFW parameterization  (section 5.2). The grey curve shows the X-ray
gas mass profile. (The best-fit X-ray gas mass profile and $1\sigma$ 
errors are plotted, although these curves are indistinguishable in the 
figure.)}
\label{fig:integrated}
\end{figure}

Fig.~\ref{fig:contour} shows a contour plot of the $\chi^2$ values
obtained for the full range of NFW models studied. The minimum
$\chi^2$ value  is marked with a cross. Contours have been drawn at
intervals of  $\Delta\chi^2=$ 2.30, 6.17 and 11.8, corresponding to
formal confidence  limits of 68.3 (1\,$\sigma$), 95.4 (2\,$\sigma$)
and 99.73 (3\,$\sigma$) per cent for two interesting
parameters. Fig.~\ref{fig:integrated} shows the best-fitting NFW mass
profile,  which has a  virial radius $r_{\rm 200} = c r_{\rm s} =
2.35^{+0.49}_{-0.33}$ Mpc and an integrated mass within the virial
radius, $M_{\rm 200} = 2.29^{+1.74}_{-0.82}  \times 10^{15}$\Msun.

\subsection{Other parameterized mass models}

We have also examined a variety of other popular  parameterized mass
models. Firstly, the model of  Moore \etal (1998):

\begin{equation}
\rho(r) = {{\rho_{\rm crit(z)} \delta_{\rm c_m}} \over 
{  ({r/r_{\rm s}})^{1.5} \left(1+{r/r_{\rm s}} \right)^{1.5}}},
\end{equation}

\noindent where 

\begin{equation}
\delta_{\rm c_m} = {100 } {c_m^3 \over  {{\rm ln}(1+{c_m^{1.5}}})}.
\end{equation}

For this model we obtain $\chi^2_{\rm min} = 8.6$ (6 DOF), with
best-fit  values for the scale radius $r_{\rm s}=5.0$ Mpc (the maximum
allowed value in our grid), concentration parameter  $c_m = 0.69 $ and
effective velocity dispersion  $\sigma=2125$\kmps (where $\sigma =
\sqrt{50} H_0(z) r_{\rm s} c_m$).

Secondly, we examined a simple, non-singular  isothermal sphere model:

\begin{equation}
\rho(r)=\frac{\sigma_{\rm iso}^2}{2\,\pi\,{\rm
G}}\,\frac{1}{r^2+r_{\rm c}^2},
\label{isoc}
\end{equation}

\noindent for which we obtain $\chi^2_{\rm min} = 4.9$ (6 DOF) with $r_{\rm
c}=45\pm10$ kpc  and $\sigma_{\rm iso}=1590\pm150$ \kmps.

We thus find that all of the two-parameter mass models  described
above provide acceptable descriptions of the  Chandra data for
RXJ1347.5-1145. We note, however, that a  singular isothermal sphere
[$\rho(r) \propto r^{-2}$] does  not provide an acceptable fit
($\chi^2_{\rm min} = 33.1$ for  7 DOF).

\section{Comparison with gravitational lensing results}

\subsection{Weak lensing}

\begin{figure*}
\vspace{0.5cm}
\hbox{
\hspace{1.0cm}\psfig{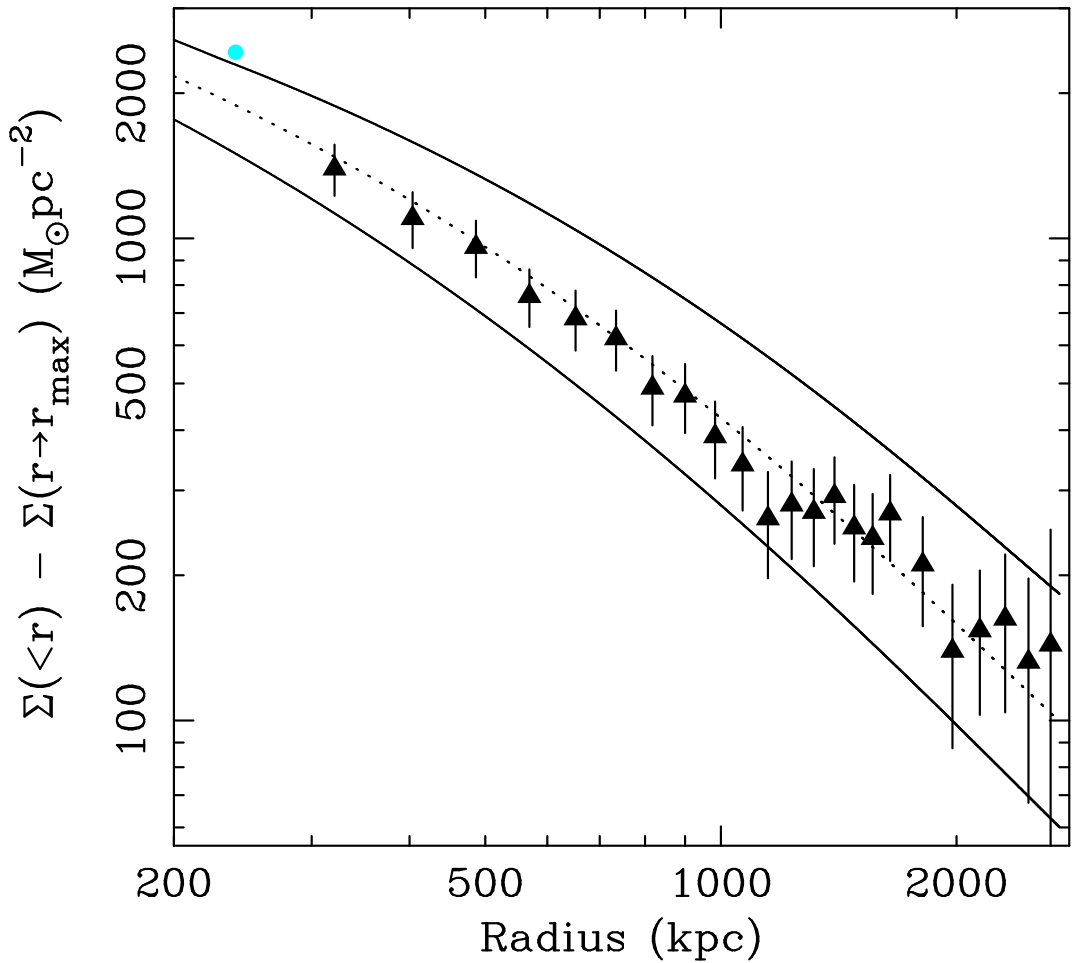}}
\caption{A comparison of the projected surface mass density contrast
determined from the Chandra X-ray data  (Section 5) with the weak
lensing results of Fischer \& Tyson (1997; solid  triangles) and the
strong lensing result from Section 6.2 (grey circle).  The best-fit
NFW X-ray mass model for the relaxed regions of the cluster  and
$1\sigma$ confidence limits (the maximum and minimum values at each
radius  for all NFW models within the 68 per cent confidence contour
shown in  Fig.~\ref{fig:contour}) are shown as the  dotted and full
curves, respectively. The strong lensing point  is marked with a grey
circle. Note that the strong lensing point should lie  above the X-ray
results, which exclude the regions of the cluster  affected by the
second mass clump. We adopt $r_{\rm max}=2.72$ Mpc as  in Fischer \&
Tyson (1997). }
\label{fig:weak}
\end{figure*}

Fischer \& Tyson (1997) present a detailed weak lensing study of 
RXJ1347.5-1145. They measure a mass profile that can be parameterized 
by an NFW model with $r_{200} = 2.4$ Mpc, or a singular isothermal
sphere with $\sigma = 1500$\kmps, in good agreement with the 
Chandra results. 

The azimuthally-averaged, projected surface mass density profile 
for RXJ1347.5-1145 determined by Fischer \& Tyson (1997) is shown in 
Fig.~\ref{fig:weak}. We have excluded the innermost point 
from their data, which extends into the strong lensing regime 
(see below; $r_{\rm max} = 2.72$ Mpc in our adopted cosmology). 
The same quantity, calculated for all NFW 
mass models within the 68 per cent confidence contour determined 
from the Chandra X-ray data (Fig.~\ref{fig:contour}) has been
 overlaid. We see that the Chandra and weak-lensing results 
exhibit good agreement on all spatial scales studied. 

The agreement between the independent lensing and X-ray mass measurements 
confirms the validity of the hydrostatic assumption used in the X-ray 
analysis (having excluded the southeast quadrant) and suggests that the mass 
profile in the cluster has been robustly determined.

\subsection{Strong lensing}

\begin{figure*}
\hspace{0.4cm} \hbox{
\psfig{figure=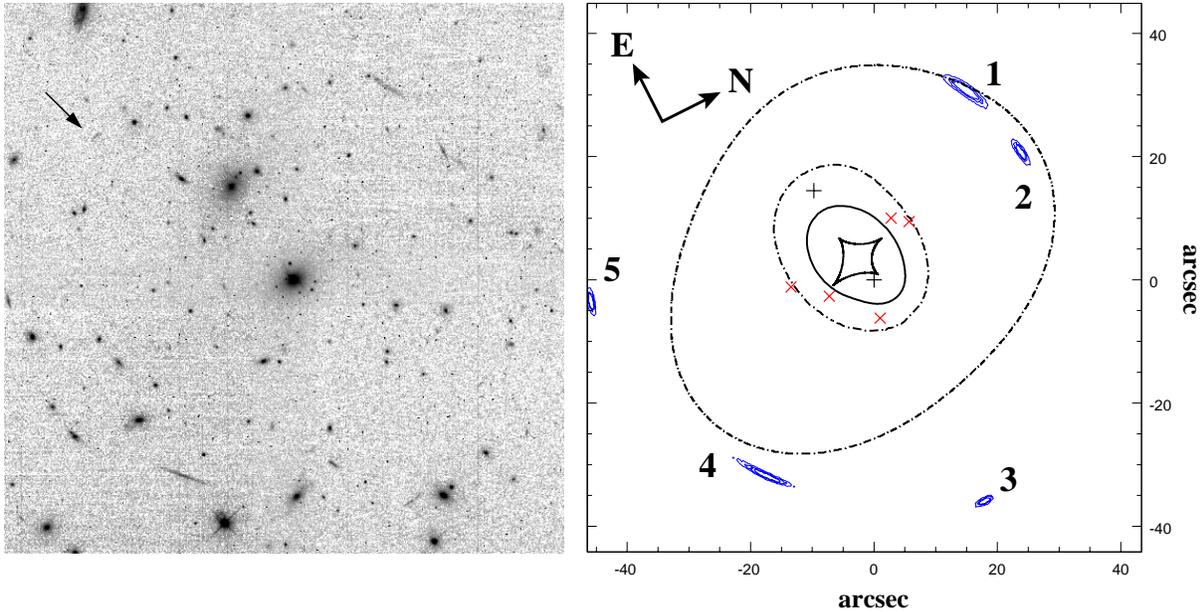,width=0.9\textwidth} }
\caption{The (a) observed and (b) predicted gravitational arc geometry 
in RXJ1347.5-1145. The image in the left panel
has been compiled from archival HST STIS data that
were originally presented by Sahu \etal (1998). The arrow points to an
additional arc-like feature mentioned in the text. The right panel shows
the arc geometry predicted by our two-component mass model. The arc numbering
has been taken from Sahu \etal (1998). The arcs have been simulated
using circular Gaussian surface brightness distributions. The 10\%,
30\% and 50\% brightness contours are plotted. The true source
positions are denoted by crosses, with details in
Table~\protect\ref{arcredshifts}. The central positions for the two
mass components are marked with plus signs.  Also shown are the
critical curves (dash-dotted) and caustic lines (solid; e.g.,
Schneider, Ehlers \& Falco 1992) for a source at a redshift of $z=0.97$.}
\label{rxjmodel}
\end{figure*}

Schindler \etal (1995) report the discovery of two bright
gravitationally-lensed arcs in RXJ1347.5-1145 using ground based
optical imaging. Sahu \etal (1998) present Hubble Space  Telescope
Imaging Spectrograph (STIS) observations of the cluster which reveal a
number of additional strongly-lensed features. The STIS image for
RXJ1347.5-1145 is shown in the left panel of Fig.~\ref{rxjmodel}. Sahu
\etal (1998) also present ground-based spectroscopy of the bright
northern arc in the cluster (arc 1) at a radius of $34.9$ arcsec (240
kpc) from the dominant cluster galaxy, for which they measure a
redshift $z_{1} = 0.806$.

Following the methods outlined in Schmidt \etal (2001), we have
examined the constraints that the observed strong lensing
configuration can place on the mass distribution in the cluster core.
We first examined a simple, spherically-symmetric mass model, centred on 
the dominant cluster galaxy, with the best-fit parameter
values determined from the Chandra data for the relaxed regions  of
the cluster (Sect.~\ref{nfwmodels}). We find that such a mass model
does not  provide a sufficient central mass density to explain the
strong lensing data.

We next examined refined lensing models in which a second mass
component, centered on the second brightest galaxy 
was introduced, together with appropriate  ellipticities
for the matter distributions. (The introduction of such a mass
clump is well-motivated by the Chandra data for the southeast 
quadrant of the cluster. The second brightest galaxy is 
likely to have been the dominant galaxy of the recently-merged 
subcluster; Section 8). Both mass components have been modelled
by elliptical NFW models (see  Schmidt \etal 2001 for
definitions). The parameters for the main  mass component were fixed
to the values determined from the Chandra data for the relaxed regions
of the cluster in Sect.~\ref{nfwmodels}, \ie  scale radius
$r_{\rm s} = 0.4$ Mpc, velocity dispersion  $\sigma=1450$\kmps~and
concentration parameter $c=5.87$. The axis ratio and position angle
were fixed to the  values measured for the dominant galaxy from  the
STIS data: axis ratio $q=0.76$ and position angle
$\theta=13.1^{\circ}$.  The centre of the main mass component was
fixed to the centre of the dominant galaxy.

The parameters for the second mass component have been determined by
requiring that the overall potential is able to produce the northern
arc (arc 1) at the measured redshift, $z_1=0.806$. For an assumed scale
radius $r_{\rm s, clump} = 0.25$ Mpc, we obtain a best-fitting
velocity dispersion $\sigma_{\rm clump} =815$\kmps~and a concentration
parameter $c_{\rm clump}=5.3$. Such parameters are typical for
mid-temperature galaxy clusters ($kT \sim$ 5 keV. \eg 
Allen \etal 2001c). The position, 
ellipticity and position angle of the second mass component
were fixed to the values determined for the second brightest galaxy
from the STIS image: $q_{\rm clump}=0.79$ and $\theta_{\rm 
clump}=28.1^{\circ}$.

The redshift measurement for the northern arc constrains the 
projected mass within the arc radius. It is possible to vary the 
scale radius for the second mass component and to adjust the concentration 
parameter accordingly, so long as the projected mass within the arc 
radius remains constant. The absence of multiple arc images 
within the cluster core suggests $r_{\rm s, clump} > 0.1$ Mpc.

\begin{table}
\begin{center}
\caption{The redshifts, sizes (in arcsec) and positions (offsets
in arcsec with respect
to the dominant cluster galaxy in the STIS reference frame) for the 
sources in the lens model shown in Fig.~\protect\ref{rxjmodel}.}
\label{arcredshifts}
\begin{tabular}{ccccc}
arc & redshift & \multicolumn{1}{c}{size} &
\multicolumn{2}{c}{position}\\
& & \multicolumn{1}{c}{(FWHM)} & $\Delta$ x & $\Delta y$\\
\hline
1 & 0.806 & 0.8 & 2.77 & 10.03\\
2 & 0.75 & 0.6 & 5.71 & 9.44\\
3 & 0.97 & 0.6 & 1.00 & -6.23\\
4 & 0.97 & 0.5 & -7.24 & -2.66\\
5 & 0.97 & 0.6 & -13.44 & -1.16
\end{tabular}
\end{center}
\end{table}

Having adjusted the parameters for the second mass component to
produce the northern arc, we then examined whether the resulting mass 
model can explain the other arcs observed by Sahu \etal (1998). 
Unfortunately, there are no measured redshifts for the other arcs. 
(Sahu \etal 1998 note a possible lower limit for arc 4 of
$z\geq1$ based on the non-detection of [O{\sc ii}] in  their
ground-based spectra, although they caution that the arc may  simply
be a galaxy without a strong emission-line feature). The details  of
our model predictions are shown in the right panel of
Fig.~\ref{rxjmodel}. The results on the predicted arc redshifts and
the  assumed sizes of the sources are listed in
Table~\ref{arcredshifts}. The predicted redshifts are cosmology
dependent and would be modified slightly if, for example, an
accelerating universe with a different Hubble Constant were
assumed. For the mass model used here, the predicted redshifts should
be accurate to $\sim\pm0.05$, with the exception of arc 4 which is
determined more precisely.  We have modelled the lensed galaxies using
circular Gaussian brightness distributions with the  full-width-at
half-maximum (FWHM) values listed in Table~\ref{arcredshifts}.  The
FWHMs for the arcs were adjusted so that their 30 per  cent brightness
contours provide a good fit to the STIS data.  The effects of using
different  source sizes is illustrated by the different brightness
contours in Fig.~\ref{rxjmodel}.

Although we include arc 2 in our modelling, we caution that this
source  appears quite red in the colour image published by Fischer \&
Tyson (1997) and that it is possible that the source is simply an
edge-on spiral galaxy. The predicted redshift for arc 2 is lower than
$z_1$, since otherwise the lens model would predict a much stronger
distortion. Arcs 3, 4 and 5 are consistent with being at the same
redshift. The projected separation of the sources producing arcs 4 and
5 is 43.4 $h_{50}^{-1}$ kpc.  Although arcs 4 and 5 have a similar
colour in the  data of Fischer \& Tyson (1997), it is not possible to
explain them as images of the same source using our mass model. In
exploring different redshifts for arc 3, we found that for redshifts
higher than $z_3\sim1$ it is possible to produce a counter image on
the other side of the central galaxy. This alerted us to the presence
of a blue object (identified from the colour image of Fischer \& Tyson
1997) to the south-east of the merging subclump (see the arrow in the
left panel of Fig.~\ref{rxjmodel}). However, both the appearance in
the STIS image and our lensing analysis suggest that these objects are
not images of the same source.

We note that circular NFW models for one or both of the main mass
components cannot explain the observed arc shapes and  orientations well. In
particular, the orientation of arc 4 reflects the orientation of the
critical curve (Fig.~\ref{rxjmodel}) in its vicinity. A small $\sim
5^\circ$ difference between the orientation of the observed and predicted
arc 4 remains in our best-fit model, which points to a small imprecision 
of the lensing model.

The grey circle in Fig.~\ref{fig:weak} shows the
azimuthally-averaged, surface mass density contrast  at $r=240$ kpc
(centred on the main mass component)  for the best-fit strong lensing
model. Note that the strong  lensing results should lie above the
X-ray results at the same radii,  since the Chandra analysis excludes
the regions of the cluster  affected by the second mass subclump. The
surface mass density  contrast rises sharply between the innermost
weak lensing point  and the strong lensing result, suggesting that the
second mass  component is centrally concentrated and does not make a
major contribution  to the total mass beyond the strong lensing
regime.

In conclusion we find that a simple two-component mass model, with
ellipticities and orientations for the mass components matching 
those of the dominant cluster galaxies, provides a reasonable 
description for the overall arc geometry in RXJ1347.5-1145. Such 
a model is consistent with the Chandra X-ray and weak lensing 
results.

\begin{figure*}
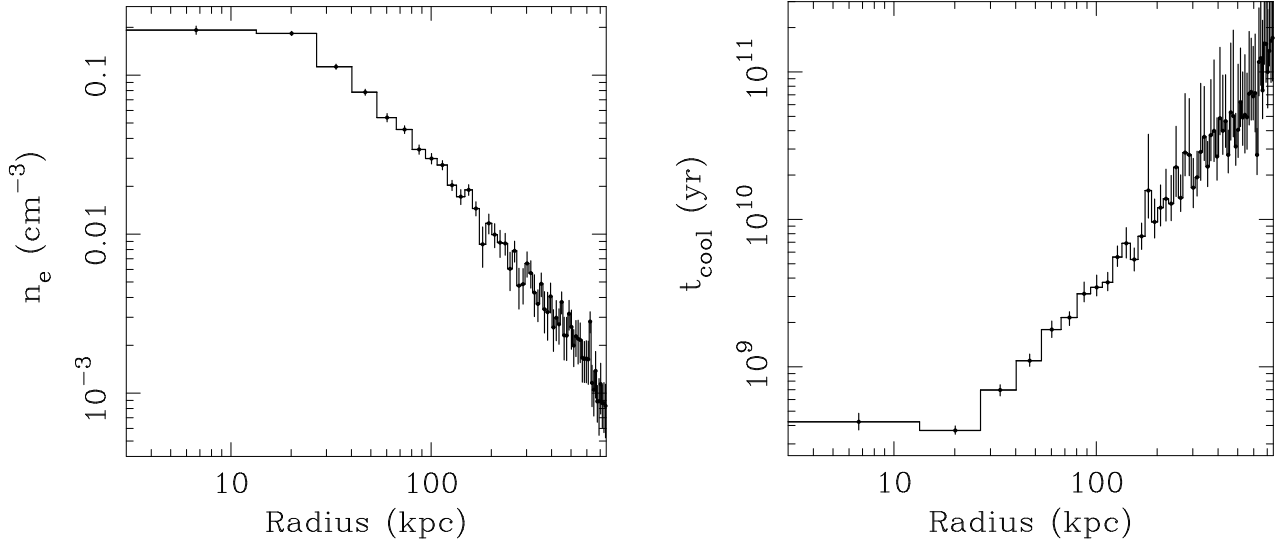

\vspace{0.5cm}
\hbox{
\hspace{0.2cm}\psfig{figure=fig10a.ps,width=0.45\textwidth,angle=270}
\hspace{0.8cm}\psfig{figure=fig10b.ps,width=0.45\textwidth,angle=270}
}
\vspace{0.3cm}
\caption{The results on (a) the electron density and (b) the cooling
time,  determined from the X-ray image deprojection analysis using the
best-fit NFW mass model. Error bars are the  $1\sigma$ errors
determined from 100 Monte Carlo simulations. A Galactic  column
density of $4.85 \times 10^{20}$ \apc~and a metallicity of 0.4 solar
are assumed.}\label{fig:deproj}
\end{figure*}

\section{The properties of the cluster gas}

\subsection{Electron density and cooling time profiles}

The results on the electron density and cooling time as a function  of
radius, determined from the image deprojection analysis using the
best-fit NFW mass model, are shown in Fig. \ref{fig:deproj}. Within
the  central 500 kpc radius, the electron density profile can be
parameterized  ($\chi^2 = 43.2$ for 34 degrees of freedom) by a
$\beta$-profile with a  core radius, $r_c = 27.8\pm1.6$ kpc,
$\beta=0.521\pm0.012$  and a central density, $n_{\rm e}(0)=2.3\pm0.1
\times10^{-2}$  cm$^{-3}$ ($1\sigma$ errors). We  measure a central
cooling time ({\it i.e.} the mean cooling time within  the central
1.97 arcsec ($\sim 13.4$ kpc) bin of  $t_{\rm
cool}=4.1^{+0.6}_{-0.5} \times 10^8$ yr (uncertainties are the 
10 and 90  percentile values
from 100  Monte Carlo simulations).

\subsection{The X-ray gas mass fraction}

\begin{figure}
\vspace{0.5cm}
\hbox{
\hspace{0.0cm}\psfig{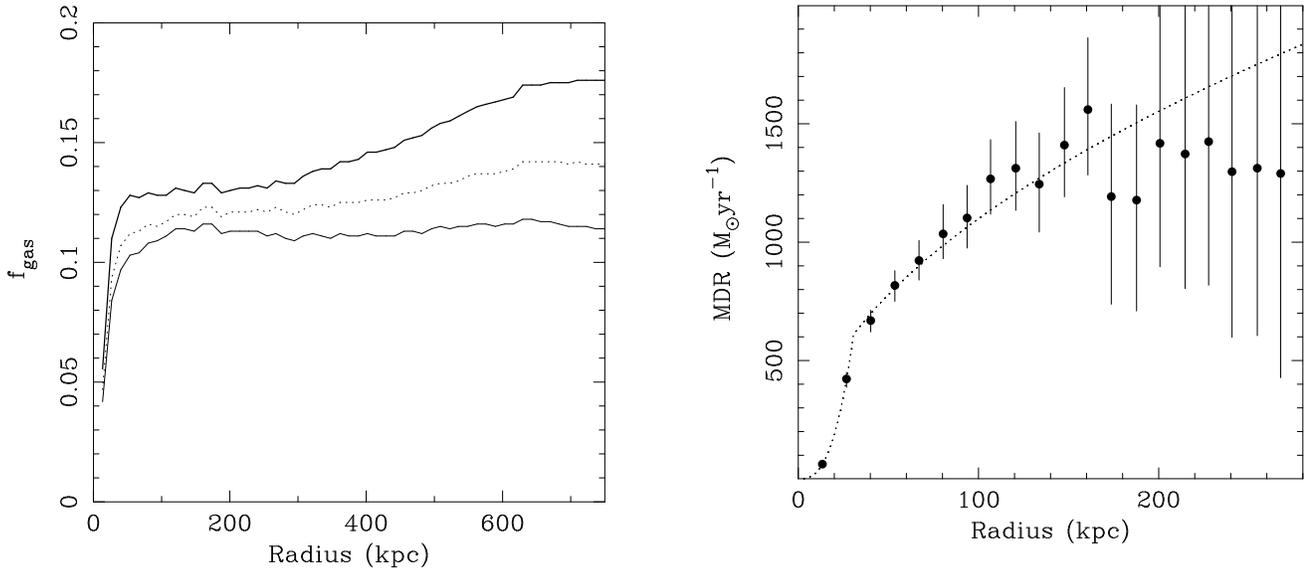}}
\caption{The ratio of the X-ray gas mass to total gravitating mass as
a function of radius. The three curves show the best-fit value (dotted
curve) and  $1\sigma$ confidence limits (solid curves).  At $r = 0.74$
Mpc we measure $f_{\rm
gas}=0.141^{+0.035}_{-0.027}$.}\label{fig:baryon}
\end{figure}

The X-ray gas-to-total-mass ratio as a function of radius, $f_{\rm
gas}(r)$,  determined from the Chandra data is shown in
Fig. \ref{fig:baryon}. We find  that the best-fit $f_{\rm gas}$ value
rises rapidly with increasing radius  within the central $r \approxlt
40$ kpc radius and then remains  approximately constant, or rises
slowly, out to the limits of the  data at $r = 0.74$ Mpc where we
measure $f_{\rm gas}=0.141^{+0.035}_{-0.027}$.

Following the usual arguments, which assume that the properties of
clusters provide a fair sample of those of the Universe as a whole
(\eg White \& Frenk 1991; White \etal 1993; White \& Fabian 1995; 
Evrard 1997; Ettori \& Fabian 1999), we may use our results on the X-ray 
gas mass fraction to
estimate the total matter density in the Universe, $\Omega_{\rm m}$.
Assuming that the luminous baryonic mass in galaxies in RXJ1347-1145
is $0.134\,h_{50}^{0.5}$ of the X-ray gas mass (\eg White \etal 1993;
Fukugita, Hogan \& Peebles 1998) and neglecting other possible sources
of baryonic dark matter in the cluster, one can show that
$\Omega_{\rm m} =  \Omega_{\rm b}/1.134 f_{\rm gas}$, where
$\Omega_{\rm b}$ is mean baryon density in the Universe.  For
$\Omega_{\rm b}h_{50}^{2} = 0.0820\pm0.0072$ (O'Meara \etal 2001) we
obtain $\Omega_{\rm m} = 0.51 \pm 0.12$. For a $\Lambda$CDM cosmology
with $H_0=70$\kmpspMpc, our results for RXJ1347.5-1145 would imply
$\Omega_{\rm m} = 0.33\pm0.08$.

\subsection{Cooling flow models}

\begin{figure}
\vspace{0.5cm} 
\hbox{
\hspace{0.0cm}\psfig{figure=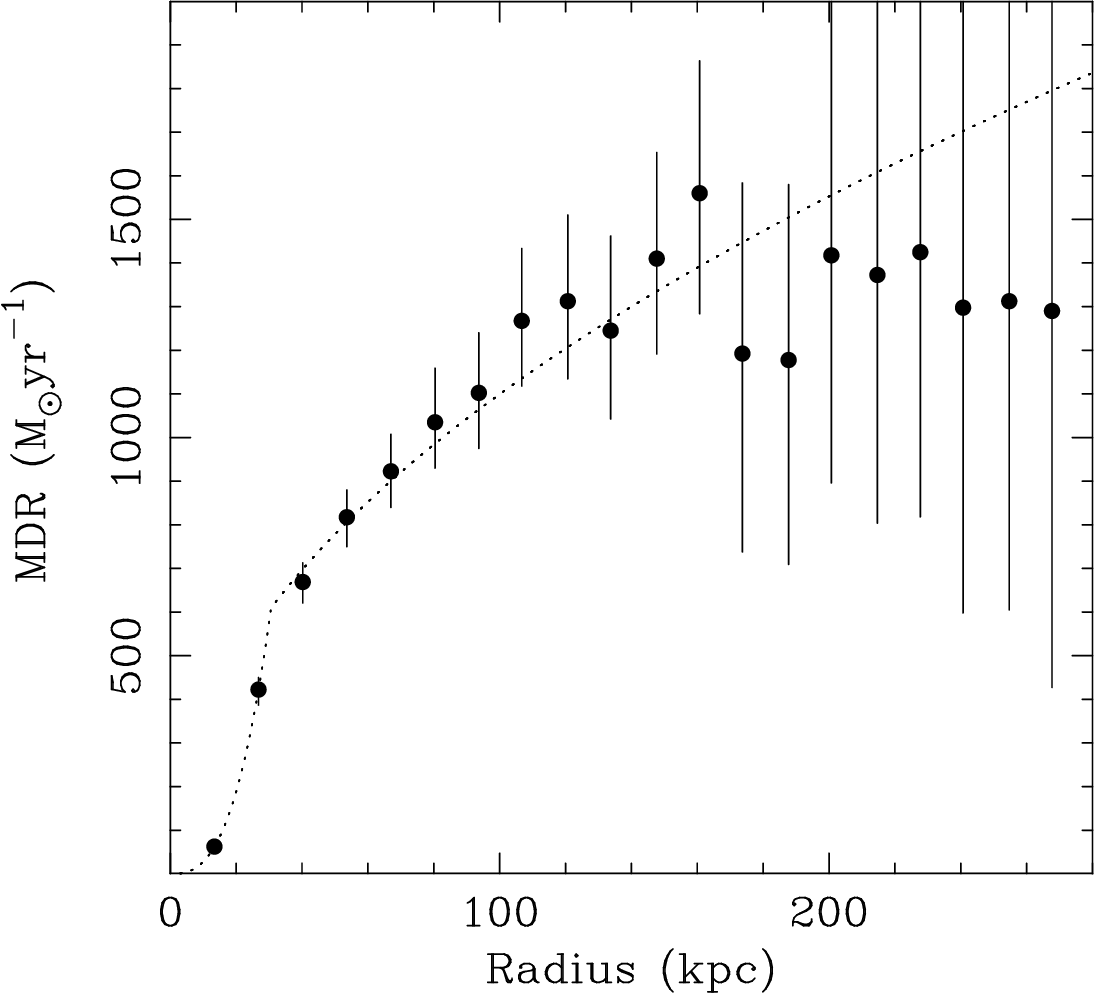,width=0.45 \textwidth,angle=270}}
\caption{The mass deposition rate (MDR) determined from  the image
deprojection analysis under the assumption that the central regions of
the cluster contain a steady-state, inhomogeneous cooling flow.  Error
bars are the 10 and 90 percentile  values from 100 Monte Carlo
simulations. The dotted line shows the best-fitting broken power-law
model described in Section 7.3.}\label{fig:mdot}
\end{figure}

Under the assumption that the X-ray gas in the core  of RXJ1347.5-1145
is in a steady-state cooling flow,  we can parameterize the luminosity
profile in terms  of an equivalent mass deposition rate (\eg White
\etal 1997).  The results of this calculation are shown in
Fig. \ref{fig:mdot}.  We see that the integrated mass deposition rate
(MDR)  rises with increasing radius within the central 100 kpc, to a
maximum value of $\sim 1300$ \Msunpyr.

We have examined the spectrum for the central 100 kpc (radius)  region
using a model appropriate for a cooling flow with distributed  mass
deposition. (We use the model of Nulsen 1998 and assume that  the
integrated MDR within radius $r$, ${\dot M} \propto r$.)  The
normalization of the cooling-flow component is  parameterized in
terms of a mass deposition rate, ${\dot M}$.  The cooling flow was
also allowed to be absorbed by  an intrinsic column density, $\Delta
N_{\rm H}$, of cold gas (with solar metallicity) located at the
redshift of the cluster.  Both ${\dot M}$ and $\Delta N_{\rm H}$ were
free parameters in the  fits. Accounting for projection effects, we
find that the  $0.5-7.0$ keV spectrum for the central 100 kpc radius
region  is consistent with a cooling flow of $\approxlt 600$ \Msunpyr,
intrinsically absorbed by a column density of $1-4 \times
10^{21}$\apc, with a mean mass-weighted flow temperature of
$\sim 10$ keV.

Fitting the mass deposition profile in Fig. \ref{fig:mdot}~with a
broken  power-law model, we determine a break radius of $31\pm2$ kpc,
and slopes internal and external to the break radius of
$2.7\pm0.3$ and $0.5\pm0.1$, respectively ($1\sigma$ errors).  Allen
\etal (2001a) argue that such breaks may mark the  outer edges of
the present-day cooling flows in clusters.  The predicted MDR
for the central $r \sim 20$ kpc radius, where a steady cooling 
flow with gas cooling to zero degrees might be expected to occur, 
is $\sim 200$ \Msunpyr. The cooling time of the X-ray gas at the break 
radius is $\sim 6 \times 10^8$ yr.

\section{Discussion}

As discussed in Section 4.2, the region of enhanced X-ray emission
(and enhanced SZ effect; Komatsu \etal 2001) to the southeast of the
main X-ray peak  is significantly hotter ($kT = 18\pm3$ keV) than at
the same distance in  other directions  ($kT = 12\pm1$ keV). Following
Markevitch \etal (1999),  we can  model the properties of this region
as a one-dimensional shock.   In this case, the  relative velocity,
$v$, of the infalling subcluster and the shock compression factor,
$\rho_1/\rho_0$, can be estimated from  the pre-shock ($kT_0$) and
post-shock ($kT_1$) temperatures. For $kT_0=12$ keV and $kT_1=18$ keV,
we obtain $v=2250$ \kmps~and $\rho_1/\rho_0 = 1.7$. If we instead
adopt $kT_0=5$ keV,  a temperature consistent with the predicted mass
of the merging subcluster from the strong lensing analysis discussed
in Section 6.2 (using the mass-temperature relation of Allen \etal
2001c), then for $kT_1=18$ keV we obtain  $v=4550$ \kmps~and
$\rho_1/\rho_0 = 3.0$. We note that the roughly spherical appearance
of the shocked gas suggests that this material has  expanded following
the initial shock. Since the material will cool as it expands, it is
likely that the initial post-shock temperature exceeded 18 keV and,
for $kT_0=5$, that the relative collision velocity exceeded 5000\kmps.
Such values are broadly consistent with the peak X-ray surface
brightness  enhancement of a factor $\sim 3$ observed at a radius of
$\sim 180$ kpc.

The second brightest galaxy in the cluster lies approximately  18
arcsec to the east and 2 arcsec to the south of the dominant  cluster
galaxy. This galaxy has an extended, diffuse halo and appears to lie at
the centre of a significant sub-concentration of galaxies within the
cluster (Fig.~\ref{rxjmodel}a). These properties are consistent  with
it having been the dominant galaxy of the recently merged
subcluster. The region of shocked X-ray gas lies approximately 11
arcsec to the south and 6.5 arcsec to the west  of the second
brightest galaxy. Assuming that the dark  matter and galaxies
associated with the subcluster have moved on ahead of the shock, 
it seems plausible that the subcluster travelled in from a  
PA of $\sim 210$ degrees. This is consistent with the  extension of the X-ray
emission observed on large scales to the south (Fig.~\ref{fig:im1}b)
and the weak lensing mass map presented by  Fischer \& Tyson
(1997). Given the separation of the shocked gas and second brightest
galaxy, we estimate that the shock probably occurred a few $10^7$
years ago. This timescale is consistent with the spatial extent of the
shocked gas, assuming that it has expanded at the sound speed, $c_{\rm
s} \sim 1500 (T/10^8{\rm K})^{-1/2}$\kmps. Note also that this
timescale is comparable  to the time required for the electrons and
ions to reach equipartition, so  it is possible that the electron and
ion populations in the shocked gas  have slightly different
temperatures.

Despite the evidence for shocked gas in the southeast quadrant, the
Chandra data for the rest of the cluster place RXJ1347.5-1145 
on the $M_{2500}-kT_{2500}$ and $kT_{2500}-L_{2500}$ relations for
relaxed, massive clusters discussed by Allen, Schmidt \& Fabian
(2001).  As discussed in Section 4.4, the shocked gas produces  only a
small rise in the mean, emission-weighted temperature  measured within
the central 0.74 Mpc radius. The excess bolometric  luminosity of the
southeast quadrant between radii of 60-195 kpc  (where the shock is
most apparent) is $\sim 10^{45}$ \ergps, which  corresponds to only
$\sim 5$ per cent of the total cluster luminosity  (Section 4.4). It
therefore appears that the overall temperature and luminosity of the
cluster have not been boosted substantially by the merger event. 
This is supported by the consistent X-ray and weak lensing mass 
results discussed in Section 6.1.

Ricker \&  Sarazin (2001) and Ritchie \& Thomas (2002) discuss
simulations of mergers between clusters with similar central densities
and argue that such mergers can lead to significant short-term
increases in the overall X-ray temperatures and luminosities of
clusters during the periods of closest  approach.  These effects are
most pronounced and long-lived in  clusters with relatively low
central densities ($n_{\rm e} \approxlt$  few $10^{-3}$cm$^{-3}$;
Ritchie \& Thomas 2002). The fact that large boosts in the overall
temperature and luminosity are not observed in RXJ1347.5-1145 may be
related to the very high central density in the main cluster core
($n_{\rm e} >  10^{-1}$cm$^{-3}$; Fig.~\ref{fig:deproj}a)  and  could
indicate that the recently-merged  subcluster had a relatively low
central gas density and interacted only weakly with the main cluster
core.

Recently, Cohen \& Kneib (2002) have published spectroscopic redshift 
measurements for 47 cluster members for which they determine a 
velocity dispersion of $910\pm130$\kmps. Inspection of their Fig. 2 
suggests that a significant fraction of the galaxies with measured redshifts 
may have been associated with the recently merged subcluster. This could 
explain the relatively low velocity dispersion, in comparison 
to the X-ray and lensing results. 

The agreement between the Chandra X-ray and gravitational  lensing
mass measurements for RXJ1347.5-1145 reinforces the  results and
conclusions drawn from previous studies of other relaxed lensing
clusters \eg Abell 2390 (Allen \etal 2001b) and  Abell 1835 (Schmidt
\etal 2001); see also the results for  MS1358.4+6245 by Arabadjis \etal
(2002). The close agreement of the independent X-ray and lensing
masses indicates that the  mass measurements are robust and
limits the contributions from non-thermal sources of pressure
support in the X-ray gas, such as bulk and/or turbulent motions and 
magnetic fields, in the relaxed regions of the cluster.

\section{Conclusions}

The main conclusions from this work may be summarized as follows:
\vskip 0.2cm

(i) We have reported Chandra observations of RXJ1347.5-1145, the most
X-ray luminous cluster of galaxies  known. We have identified a region
of shocked gas (enhanced  X-ray brightness and temperature) to the
southeast of the  main X-ray peak, at a position consistent with the
region of enhanced SZ effect reported by  Komatsu \etal (2001). The 
shocked gas probably results from recent subcluster merger activity.
The merger appears not to have boosted the overall luminosity and 
temperature of the cluster substantially. 

(ii) Excluding the data for the southeast quadrant, we have  measured
the density, temperature and mass profiles for  the cluster. The mass
profile can be  parameterized by an NFW model with a scale radius
$r_{\rm s} = 0.40^{+0.24}_{-0.12}$ Mpc and a  concentration parameter,
$c=5.9\pm1.4$ (68 per cent confidence  limits). The normalization of
the mass profile may also  be expressed  in terms of an effective
velocity dispersion, $\sigma = \sqrt{50}  H(z)r_{\rm s} c =
1450^{+300}_{-200}$\kmps.

(iii) The best-fit Chandra mass model for RXJ1347.5-1145  is in
good agreement with independent measurements  from weak
gravitational lensing studies  (Fischer \& Tyson 1997). The observed
strong lensing  configuration in the cluster core can also be
explained with the introduction of an additional mass clump centred
on the second brightest galaxy, which  is likely to have been the
dominant galaxy of the recently merged subcluster.

\section*{Acknowledgements}

We thank R. Johnstone and S. Ettori for supplying a number of 
scripts used in the X-ray reduction/analysis and Philippe Fischer for 
communicating his weak lensing results in a machine readable form. 
SWA and ACF acknowledge the support of the Royal Society.

\end{document}